\documentclass[english,a4paper]{article}
\usepackage{mathptmx}

\usepackage[T1]{fontenc}
\usepackage[latin9]{inputenc}
\usepackage{float}
\usepackage{graphicx}

\makeatletter

\newcommand{\noun}[1]{\textsc{#1}}
\providecommand{\tabularnewline}{\\}

\newcommand{\lyxaddress}[1]{
\par {\raggedright #1
\vspace{1.4em}
\noindent\par}
}

\usepackage{mathptmx}

\usepackage{float}

\makeatletter

\usepackage{mathptmx}

\usepackage{geometry}

\usepackage{float}

\usepackage{color}

\makeatletter

\usepackage{axodraw}

\makeatother

\makeatother

\makeatother

\usepackage{babel}

\begin{document}

\title{Fourth Family Neutrinos and the Higgs Boson at the LHC}

\author{T. Çuhadar-Dönszelmann$^{1}$, M. Karagöz Ünel$^{2}$, V. E. Özcan$^{3}$,
S. Sultansoy$^{4,5}$, G. Ünel$^{6,7}$}

\maketitle

\lyxaddress{\begin{center}
$^{1}$University of Sheffield, Department of Physics and Astronomy,
Sheffield, UK\\
 $^{2}$University of Oxford, Department of Physics, Oxford, OX1 3RH,
UK\\
 $^{3}$University College London, Department of Physics and Astronomy,
London, UK\\
 $^{4}$TOBB University of Economics and Technology, Physics Department,
Ankara, Turkey\\
 $^{5}$Institute of Physics, Academy of Sciences, Baku, Azerbaijan\\
 $^{6}$CERN, Physics Department, Geneva, Switzerland\\
 $^{7}$University of California at Irvine, Department of Physics,
Irvine, CA, USA 
\par\end{center}}

\begin{abstract}
We evaluate the LHC discovery potential for the fourth family Standard
Model neutrinos in the process $pp\rightarrow Z/h\rightarrow\nu_{4}{\bar{\nu}_{4}}\rightarrow W\mu W\mu$.
We show that, depending on their masses, the simultaneous discovery
of both the Higgs boson and the heavy neutrinos is probable at early
stages of LHC operation. Results are presented for both Majorana and
Dirac type fourth family neutrinos. 
\end{abstract}

\section{Introduction}

The main goal of the LHC experiments is the vindication or rejection
of the Higgs mechanism as the underlying cause of fermion masses in
the Standard Model (SM). Higgs boson searches are therefore, of utmost
importance. Understanding the flavor structure of the SM, in particular,
determining the number of fermion families, is also a key goal. The
data from LEP-1 strongly favored three families of fermions with light
neutrinos ($m_{\nu}<m_{Z}/2$) \cite{PDG}. Thus, there are no experimental
or phenomenological evidence excluding the existence of a fourth fermion
family with a heavy neutrino. Indeed, the recent electro-weak precision
data are equally consistent with the presence of three or four fermion
families \cite{4Fam-1,4Fam-2}, whereas the four family scenario is
favored if the Higgs is heavier than 200 GeV \cite{tait}. These compelling
reasons form a primary argument to search for a fourth SM family with
heavy fermions. A secondary impetus arises from the as yet unexplained
hierarchy observed in fermion masses. If there were four SM families
and their Yukawa couplings were identical, then the diagonalization
of the 4$\times$4 mass matrix in which all elements are unity, would
yield a single non-zero element ($M_{44}$)$\,$\cite{DMM1,DMM2,DMM3}.
In this case, the observed masses of fermions in the first three families
can be obtained from perturbations on uniform 4$\times$4 mass matrices
\cite{DMM4,DMM5,DDM-Param}. This idea is referred to in the literature
as {}``flavour democracy'': see the review \cite{FLDem review}
and references therein. A third and more recent motivation for the
fourth family arises from the proposed charge-spin unification \cite{mankoc}.
Finally, recent measurements from the B factories and the Tevatron
have shown deviations from the SM, which have been attributed to the
possible existence of a fourth generation \cite{hou-1,hou-2,soni}.

From an experimentalist's point of view, a heavy quark and a heavy
neutrino are both very interesting particles to search for at the
LHC. Searches for heavy quarks of the fourth SM family have been considered
elsewhere \cite{R-atlas-tdr,FF-all,Holdom,FF-erkcan}. Heavy neutrinos
can be produced in pairs at the LHC and are expected to decay to a
$W$ boson and a charged lepton with flavor dependent on the particular
Maki-Nakagawa-Sakata (MNS) matrix \cite{DDM-Param}. The Majorana
or Dirac nature of the fermions will have an important impact on the
observed outcome: those final states, in which both leptons have the
same sign, are expected to be free of direct backgrounds and therefore
offer a distinct signature. The use of such signatures for Higgs boson
discovery via a so-called {}``silver mode'' was recently proposed
\cite{silver_mode}. While inclusive final states with single fourth
family members might be enhanced by lower production-energy threshold,
their production cross-sections have a sine-squared dependence on
mixing angle and thus, are heavily suppressed.

In this paper, the impact of fourth family quarks on the Higgs boson
production and subsequent decay into fourth family neutrinos are considered
in detail. The $Z$ boson mediated production of the heavy neutrinos
and their decay are also studied for Higgsless scenarios.

\section{Fourth Family Neutrinos at the LHC}

The 3-family SM is extended with an additional set of quarks and leptons
denoted as: $u_{4}$ and $d_{4}$ for quarks, $e_{4}$ for the charged
lepton and $\nu_{4}$ for the heavy neutrino. The fermion-boson interaction
vertices of the fourth family fermions are similar to the known first
three families. Although the masses and the mixings of the new fermions
are not fixed, the lower bound on the mass of the fourth family quarks
from Tevatron experiments is 250 GeV \cite{R-CDF-t'}. Following the
flavour democracy approach, the masses of $u_{4}$ and $d_{4}$ are
taken degenerate and represented as $m_{q_{4}}$.

\begin{figure}[H]
\begin{centering}
\includegraphics[scale=1.5]{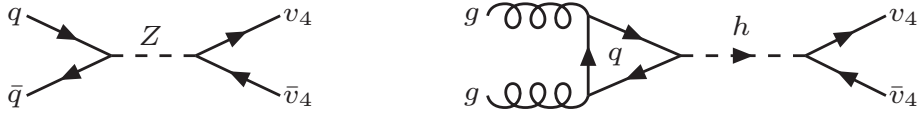} \vspace{-1.3cm}

\par\end{centering}

\caption{Possible $\nu_{4}$ pair production via Z (left) or Higgs boson (right)
at the LHC.}

\label{fig:pair_prod_feyn_diag} 
\end{figure}

The tree-level diagrams for the pair production of the fourth SM family
heavy neutrinos are shown in Figure \ref{fig:pair_prod_feyn_diag}.
The pair production cross section of the virtual $Z$ boson mediated
channel depends on the mass of the $\nu_{4}$ while that of Higgs
mediated channel depends on the Higgs and $\nu_{4}$ masses as well
as the $q_{4}$ mass, which contributes to the quark loop in Figure
\ref{fig:pair_prod_feyn_diag}.

\subsection{Impact of Fourth Family Quarks on Higgs Boson Production}

\begin{figure}[h]

\lyxaddress{\begin{centering}
\includegraphics[scale=0.4]{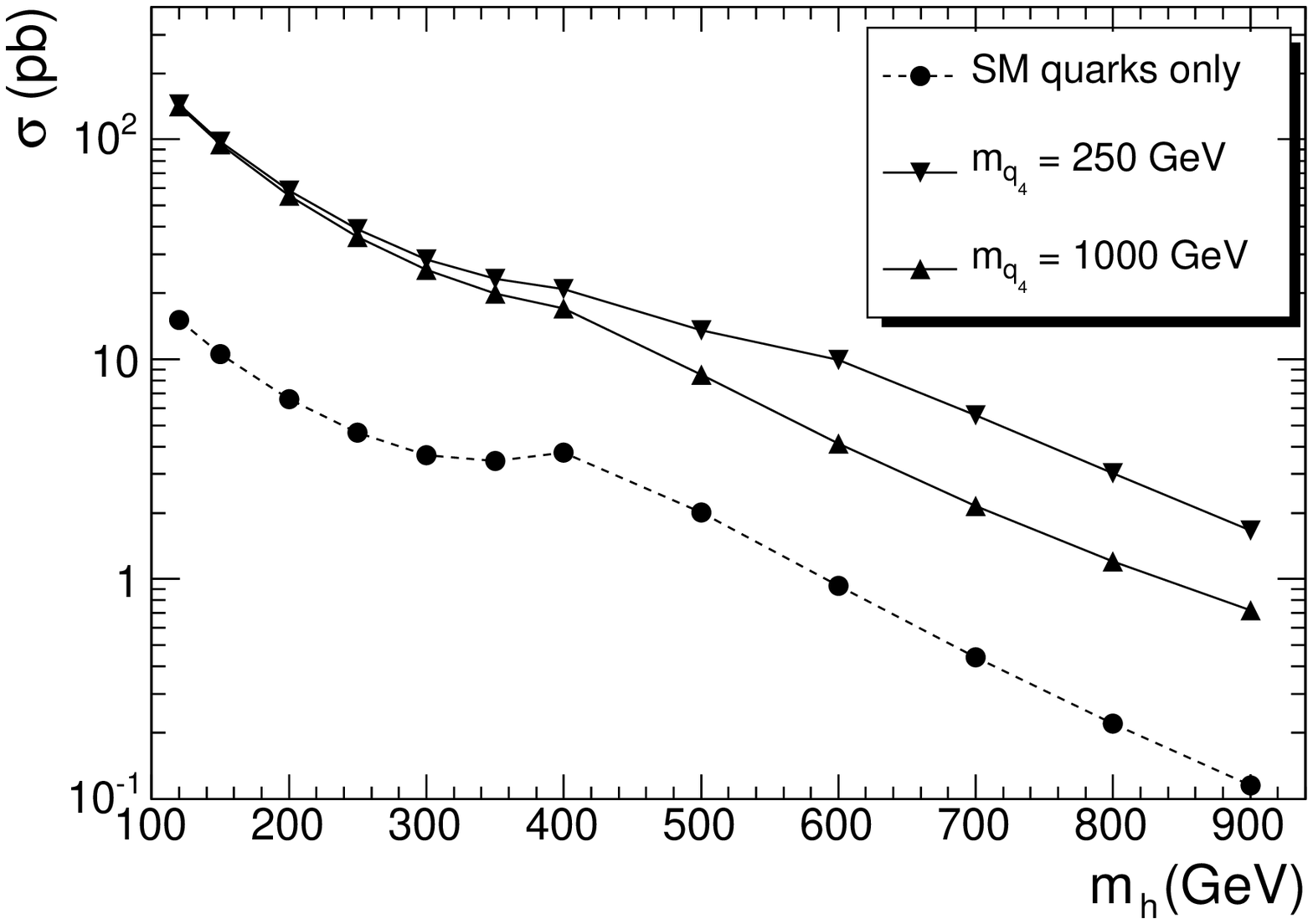} \includegraphics[scale=0.4]{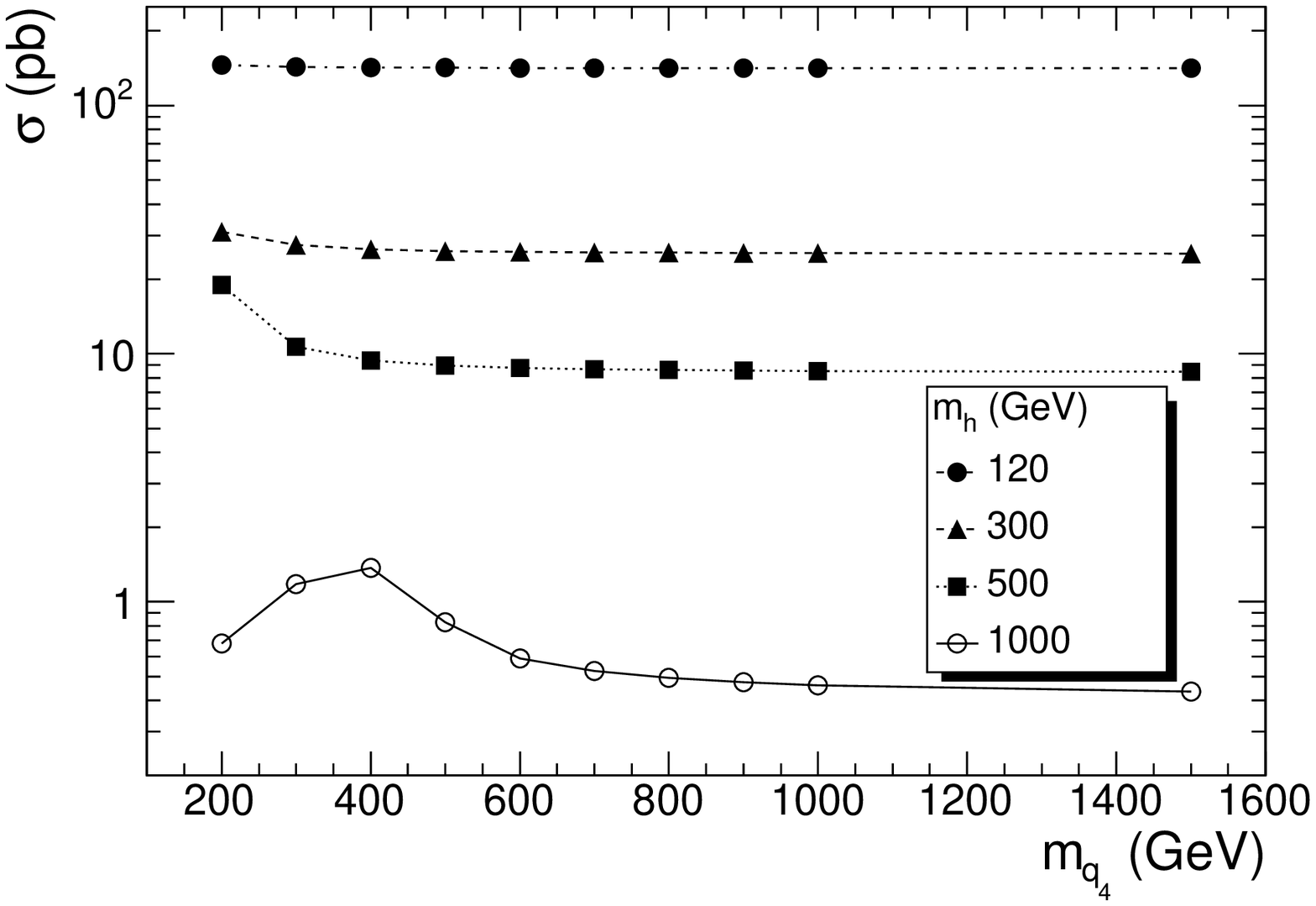} 
\par\end{centering}}

\caption{Higgs boson production cross section as a function of Higgs mass,
for SM (circles), SM + fourth family for $m_{q_{4}}=1000$$\,$GeV
(upwards triangles) and similarly for $m_{q_{4}}=$250$\,$GeV (downwards
triangles) (left). Higgs boson production cross section as a function
of the new quark mass, for different Higgs boson mass values: 120,
300, 500 and 1000$\,$GeV (right).\label{fig:higgs-prod-enh}}

\end{figure}

The enhancement of the Higgs production cross section at the LHC due
to the fourth family quarks via the gluon loop was previously calculated
with the infinite mass quark approximation \cite{ff_enhancement}.
For a more realistic cross section calculation, we modified the Higgs
production cross section software, \texttt{Higlu} \cite{Higlu}, to
include the effects of the fourth family quarks with definite masses.
In Fig.~\ref{fig:higgs-prod-enh} left side, the Higgs production
cross section of 3-family SM is compared with that of 4-family SM,
for $m_{q_{4}}=250$$\,$GeV and $m_{q_{4}}=1000$$\,$GeV. It is
seen that, by comparison to the results in \cite{ff_enhancement},
$m_{q_{4}}=$ 1000$\,$GeV is a good approximation to the infinite
mass approximation. To further investigate the validity of this approximation,
the same cross section is also plotted in Fig.~\ref{fig:higgs-prod-enh}
right side, as a function of the fourth family quark mass ($m_{q_{4}}$)
for Higgs boson mass values of 120, 300, 500 and 1000 GeV. It is seen
that for a Higgs boson of $m_{h}\leq300$GeV the production cross
section is independent of the $m_{q_{4}}$, however for $m_{h}\geq500$$\,$GeV,
the deviation in the cross section is substantial if $m_{q_{4}}=400$$\,$GeV.
Therefore, for the rest of this note, for the Higgs production cross
section values, we use the leading order (LO) results obtained with
\texttt{Higlu} in the presence of a fourth family with $m_{q_{4}}=500$$\,$GeV.

\subsection{Heavy Neutrino Discovery Channels}

\begin{figure}

\lyxaddress{\begin{centering}
\includegraphics[scale=0.4]{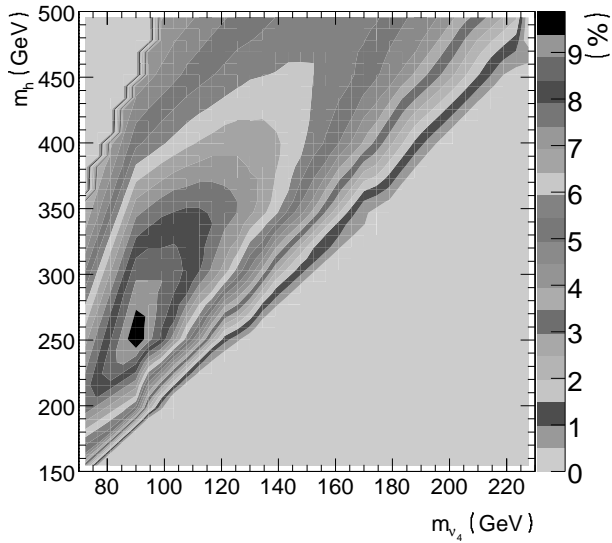} \vspace{-0.3cm}
 
\par\end{centering}}

\caption{Branching fraction of the Higgs boson decaying into $\nu_{4}$ pairs
in the 2D plane of Higgs and $\nu_{4}$ masses.\label{fig:Higgs_nu4_2dplane}}

\end{figure}

\begin{figure}

\lyxaddress{\begin{centering}
\includegraphics[scale=0.4]{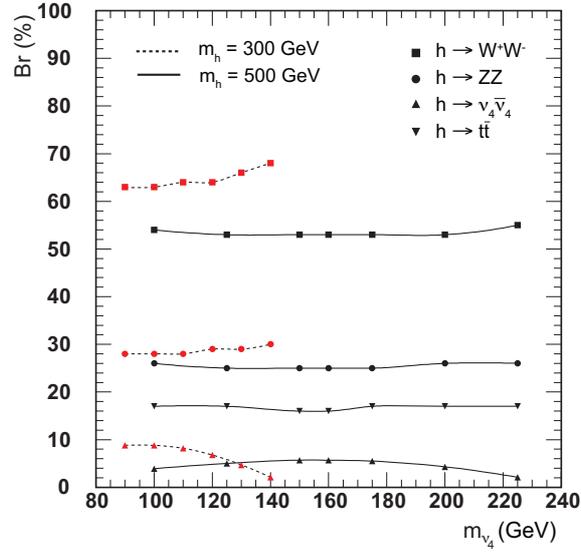} 
\par\end{centering}}

\caption{Branching fractions of the Higgs boson decaying into W, Z, $\nu_{4}$
or top-quark pairs, in the presence of a fourth family with $m_{q_{4}}=m_{e_{4}}=500$GeV.
The dashed (solid) line corresponds to the branching fraction at $m_{h}=300(500)$$\,$GeV.
\label{fig:Higgs_BR_with_nu4}}

\end{figure}

The Higgs boson branching fractions (Br) in the presence of the fourth
SM family depend on their masses. Figure \ref{fig:Higgs_nu4_2dplane}
shows the Higgs branching fraction to the fourth family neutrinos
in the $m_{h}$ vs $m_{\nu_{4}}$ plane with $m_{q_{4}}=m_{e_{4}}=500$GeV.
It is observed that the highest $Br(h\rightarrow\nu_{4}\bar{\nu}_{4})$
of $10\%$ is obtained for the values of $m_{h}=250$$\,$GeV and
$m_{\nu_{4}}=90$$\,$GeV. The branching fractions of the Higgs boson
decaying into its main channels such as $W^{+}W^{-},\, ZZ,\, t\overline{t}$
are presented in Figure \ref{fig:Higgs_BR_with_nu4} as a function
of the $\nu_{4}$ mass, for $m_{h}=300$$\,$GeV and $m_{h}=500$$\,$GeV.
The width of the Higgs boson at these two mass values is around 9
and 67 GeV respectively.

\begin{figure}

\lyxaddress{\begin{centering}
\includegraphics[scale=0.4]{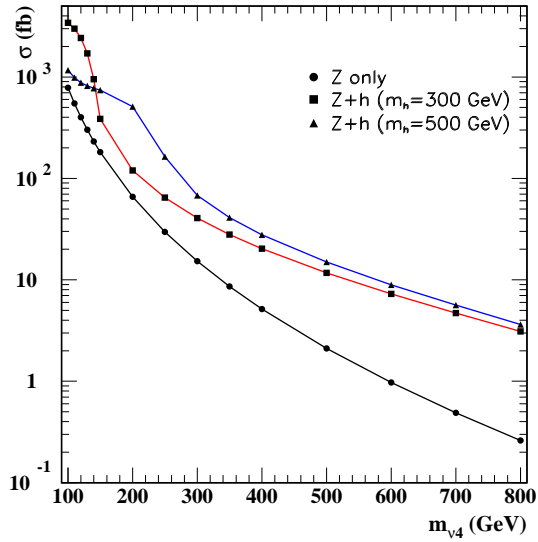} 
\par\end{centering}}

\caption{$\nu_{4}\overline{\nu}_{4}$ pair production cross section as a function
of $\nu_{4}$ mass for three scenarios: Higgsless case and cases with
Z+h ($m_{h}=300$$\,$GeV and $m_{h}=500$$\,$GeV). The enhancement
from gluon fusion is calculated for $q_{4}=$$u_{4}\,$,$\, d_{4}$
mass of 500$\,$GeV. \label{fig:prod_Z_H_overlay}}

\end{figure}

In the absence of the Higgs boson, the virtual Z boson mediated channel
will, in fact, be the only $\nu_{4}$ pair production mechanism. The
cross section of the $\nu_{4}$ pair production is calculated for
three cases: Higgsless scenario (i.e. the $Z$-boson channel), Higgs
with mass $m_{h}=300$$\,$GeV and $m_{h}=500$$\,$GeV.%
\footnote{Electroweak precision data, in the presence of some new physics, favors
high masses for the Higgs boson$\,$\cite{HeavyHiggs}.%
} The results of this calculation, as a function of $\nu_{4}$ mass
are shown in Fig. \ref{fig:prod_Z_H_overlay}. For the detailed study
of implications at the LHC, one benchmark point for each case is selected,
hereafter represented by \emph{S1}, \emph{S2} and \emph{S3}. The properties
of these benchmark points and corresponding effective cross sections
are given in Table \ref{tab:benchmark_points} for $m_{q_{4}}$=500
GeV.  The effective cross sections in the last column of Table$~$\ref{tab:benchmark_points},
with $WW\mu\mu$ fi{}nal state, are calculated using the branching
fractions given in$~$\cite{DDM-Param}. In that study, the four-dimensional
CKM matrix has been parameterized as a modification of $4\times4$
unit matrix, and the values for the three degrees of freedom in this
parameterization have been extracted from the available experimental
data. The parameterization is common between the quark and lepton
sectors and predicts $Br(\nu_{4}\to W\mu)=0.68$ for different values
of the assumed unified Yukawa coupling coefficent and the corresponding
values of the aforementioned parameters.

It is worth noting that a similar study of Higgs-mediated production
of heavy neutrinos has been performed for the Superconducting Super
Collider, but with significantly less emphasis on the background estimations$\,$\cite{SSC}.

\begin{table}[h]
\caption{Benchmark points for the $\nu_{4}$ discovery with $m_{q_{4}}$ =
500 GeV. \emph{S1} point is for the $Z$ boson mediated case, \emph{S2}
$m_{h}=300$ GeV, \emph{S3} $m_{h}=500$ GeV. The cross sections of
\emph{S2} and \emph{S3} include the contribution from the Z boson.
\label{tab:benchmark_points}}

\smallskip{}

\begin{centering}
\begin{tabular}{c|c|c|c|c|c|c}
 & $\sigma_{pp\to Z\to\nu_{4}\bar{\nu}_{4}}$ (fb)  & $m_{h}$ (GeV)  & $\sigma_{gg\to h}$ (pb)  & $m_{\nu_{4}}$ (GeV)  & BR($h\to\nu_{4}\bar{\nu}_{4}$ )  & $\sigma_{pp\to\nu_{4}\bar{\nu}_{4}\to WW\mu\mu}$ (fb)\tabularnewline
\hline
\hline 
\emph{S1}  & 782  & N/A  & N/A  & 100  & N/A  & 362\tabularnewline
\hline 
\emph{S2}  & 782  & 300  & 30  & 100  & 0.088  & 1583\tabularnewline
\hline 
\emph{S3}  & 144  & 500  & 10  & 160  & 0.055  & 321\tabularnewline
\end{tabular}
\par\end{centering}
\end{table}

\section{Analysis Strategy}

The final experimental signature depends on the nature of $\nu_{4}$.
If $\nu_{4}$ is of Majorana type, the decay products would be two
same-sign (SS) leptons and bosons half of the time and opposite-sign
leptons and bosons for the rest of the time. The case with SS leptons
has no direct SM model processes to contribute to its background.
If $\nu_{4}$ is a Dirac particle, the signature will be with two
opposite-sign leptons all the time. In brief, either same-sign or
opposite-sign high $p_{T}$ dileptons are produced in association
with two W bosons. The leptons in the event can be used for triggering.
The $W$ bosons can be reconstructed from their hadronic decays and/or
using leptonic decay of one $W$ boson to reduce the combinatoric
background due to high jet multiplicity. A full reconstruction of
an event make the measurement of the mass and the width of the both
Higgs boson and $\nu_{4}$ possible. In this paper, only the hadronic
decay channels of both $W$ bosons (thus their reconstruction) are
considered.

\subsection{Signal Properties}

\begin{figure}

\lyxaddress{\begin{centering}
\includegraphics[scale=1.4]{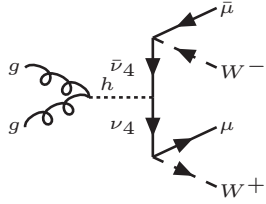}
\par\end{centering}}

\caption{Feynman diagram for the $ggh$ effective coupling vertex also showing
final states with $W$ and $\mu$. }

\lyxaddress{\label{fig:ggh-effective} }
\end{figure}

A tree level signal generator, \texttt{CompHEP} 4.4.3 was used to
implement the 4-family SM \cite{R-comphep}. We have implemented the
loop level process $gg\to h$ through an effective $ggh$ vertex coupling
into \texttt{CompHEP}. The coupling strength was adjusted to match
the LO \texttt{Higlu} results. The production of the Higgs boson via
the $ggh$ vertex and its decay via the fourth family neutrinos is
shown in Fig.~\ref{fig:ggh-effective} for opposite sign final states.

We have generated signal events using \texttt{CompHEP} and background
events using \texttt{MadGraph} 4.2.0$\,$\cite{madgraph}. The compatibility
between these two tree-level Monte Carlo generators has been previously
discussed \cite{E6-ilk}. The generated events are further processed
in \texttt{PYTHIA} 6.4.14 \cite{R-pythia} for hadronization, addition
of multiple interactions and underlying event as well as initial and
final state radiation. Finally, a fast simulation of the detector
effects, such as acceptance and resolution, is performed with \texttt{PGS}$\,$\cite{pgs}
using the parameterization for the ATLAS detector \cite{R-atlas-tdr}.%
\footnote{Since PGS does not simulate any muon mischarge, this feature was manually
added with the mischarge rate parameterized as a function of the muon
transverse momentum ( $\epsilon_{mischarge}=10^{-4+P_{T}/200{\rm {GeV}}}$
) \cite{CSC}.%
}

\subsection{Background Processes}

The main background to $\nu_{4}$ pair production is massive diboson
associated dimuon production: $2V+2\mu$ where $V=W,Z$. In the case
of a Majorana $\nu_{4}$, there are no direct SM model processes to
contribute to the background, making this channel experimentally appealing.
In the case of a Dirac $\nu_{4}$, we calculated the total direct
SM backgrounds in $2\mu+2V$ state with \texttt{MadGraph} where the
renormalization and regularization scale was set to the mass of the
Z boson and CTEQ6L1 was selected as the PDF set$\,$\cite{R-cteq}.
The breakdown of the most dominant SM background processes and their
cross sections can be found in Table \ref{bgrd_tab} left side. A
minimum $p_{T}$ requirement of 15$\,$GeV was imposed at the generator
level in \texttt{MadGraph}. It is evident from the table that the
direct SM backgrounds even for the Dirac case are essentially negligible.
The most generic formulation of the background processes is in fact
$2\mu4j$ final state. However, the computational power at hand was
not sufficient to compute the cross section and generate events with
\texttt{MadGraph}. Since the main contribution to the $2\mu\,4j$
final state would come from $\gamma/Z+4j$ events, a dedicated software,
\texttt{AlpGen} 2.1.3 \cite{AlpGen} was used to calculate their tree-level
cross section. With the previously mentioned generator level selection
criteria, the cross section is found to be 56.7$\pm$$\,$0.4 pb.
For the event generation, a shorter conservative alternative method
is applied%
\footnote{We prefer \texttt{MadGraph} because of the ease in running it on our
computational sources. With a small sample of \texttt{AlpGen}-generated
events, we have validated that our results are indeed pessimistic.%
}: all processes yielding the $\gamma/ZWjj$ final states%
\footnote{Both on-shell and off-shell $\gamma$ and $Z$ are considered.%
} are studied with \texttt{MadGraph} to calculate the cross section
(as listed in Table \ref{bgrd_tab} right side) and to generate events
which are then scaled up to the full cross section obtained from \texttt{AlpGen}
(same Table, last line). The conservativeness of the approach comes
from the fact that, in the worst case scenario, the jets in the final
state would truly come from the decay of a $W$ boson, otherwise from
an underlying event or from QCD radiation. The last two are easier
to eliminate by reconstructing the $W$ boson invariant mass. Therefore
the sole consideration of the $W$ bosons, as the source of the jets
in the final state, is a conservative approach. These events are considered
as direct background for the rest of this note.

For the indirect SM background, we consider the $t\bar{t}$ pair production
as the overwhelming candidate with a total cross section of 754.7$\,$$\pm$$\,$1.0
pb, calculated with \texttt{MadGraph}. The top quark pair production
will produce a $2W+2b_{j}$ final state, which makes it a candidate
for indirect background through misidentification and additional false
jet combinatorics for the $W$ boson reconstruction. A possible way
for such a case to fake the signal final state would be to have $W$
bosons decay leptonically with small neutrino energy in the presence
of additional jet activity (e.g. initial or final state radiation).
If both bosons decay leptonically with small neutrino energy then
the combination of a high energy lepton and a light jet or a $b$-jet
can mimic the signal. Therefore the $t\bar{t}$ pair production is
considered as the indirect background for the remainder of this note.

\begin{center}
\begin{table}
\caption{\label{bgrd_tab}SM diboson and $2\ell+4j$ backgrounds with corresponding
cross sections. All values are MadGraph results except the $\gamma/Z{}_{\to\mu^{+}\mu^{-}}\,4j$
which is obtained from AlpGen.}

\begin{centering}
\begin{tabular}{c|c}
Process  & cross section (fb)\tabularnewline
\hline
\hline 
$W^{+}W^{-}\mu^{+}\mu^{-}$  & 2.56 $\pm$ 0.02\tabularnewline
\hline 
$ZZ\mu^{+}\mu^{-}$  & 0.70 $\pm$ 0.06\tabularnewline
\hline 
$W^{+}Z\mu^{+}\mu^{-}$  & 0.97 $\pm$ 0.01\tabularnewline
\hline 
$W^{-}Z\mu^{+}\mu^{-}$  & 0.48 $\pm$ 0.06\tabularnewline
\hline
\hline 
Direct Total  & 4.71 $\pm$0.09\tabularnewline
\end{tabular}~~~~~~\begin{tabular}{c|c}
Process  & cross section (fb)\tabularnewline
\hline
\hline 
$\gamma_{\to\mu^{+}\mu^{-}}W\, jj$  & ~~~~~~80.2 $\pm$ 1.7\tabularnewline
\hline 
$Z{}_{\to\mu^{+}\mu^{-}}^{}W\, jj$  & ~~~630.1 $\pm$ 7.1\tabularnewline
\hline
\hline 
Total  & ~~~710.3 $\pm$ 7.3\tabularnewline
\hline
\hline 
$\gamma/Z{}_{\to\mu^{+}\mu^{-}}\,4j$  & 56645.4 $\pm$ 373\tabularnewline
\end{tabular}
\par\end{centering}
\end{table}

\par\end{center}

\subsection{Event Selection and Reconstruction}

\begin{figure}
\begin{centering}
\includegraphics[scale=0.65]{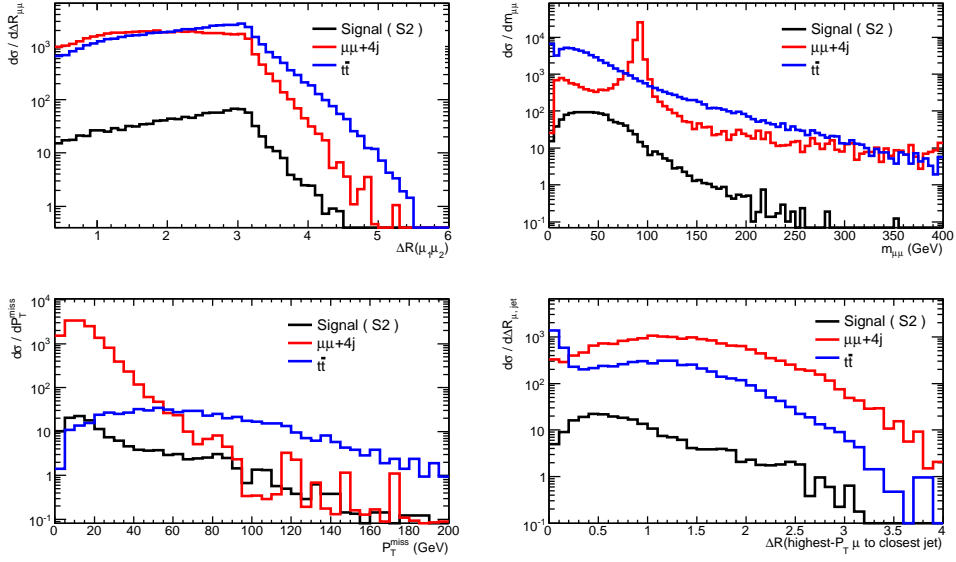} 
\par\end{centering}

\caption{Kinematic distributions for the backgrounds and the signal (benchmark
point $S2$). In all plots, the black solid lines represent the signal
events.\label{fig:S1-kinematics}}

\end{figure}

ROOT framework \cite{Root} is used to analyze the final physics objects
(such as muons and jets) provided by the simulation software. The
signal and background event samples are treated in the same analysis
code used to isolate the $\nu_{4}$ and $h$ candidates. The events
are first tagged by the existence of at least two muons with a minimum
$p_{T}$ of 15$\,$GeV. When there are more than two such muons, the
two with the highest transverse momentum are considered. As the {}``silver
mode'' analysis concentrates on the hadronic decays of the $W$ bosons
originating from the heavy neutrinos, the remaining events are required
to have at least 4 jets with a minimum $p_{T}$ of 15$\,$GeV on each
jet. All available jets are combined to find the best two $W$ boson
candidates by taking the pair with the smallest difference from the
true value of $m_{W}$ \cite{PDG}. A further selection is applied
to restrict the reconstructed invariant masses of the dijet candidates
to be within 20$\,$GeV of the $W$ boson mass. To reject muons from
the decays of the $b$ quarks, an isolation criterion is applied:
if $\Delta R$ between a muon and the closest jet of $P_{T}>20$$\,$GeV
is less than 0.4, the event is rejected. The $\Delta R_{\mu j}$ distribution
for signal and two background event types are shown in Fig.~\ref{fig:S1-kinematics}
lower left plot at the benchmark point $S2$. As the signal events
do not contain any missing energy nor any $b$-tagged jets, these
properties are used to suppress the \textbf{$t\bar{t}$} background.
The $E_{T}^{miss}$ distribution for signal and two background event
types are shown in Fig.~\ref{fig:S1-kinematics} lower right plot
at the benchmark point $S2$. The efficiencies of all the selection
criteria are listed in table \ref{tab:efficiencies}. The last row
shows the common reconstruction efficiency, $\epsilon_{reco}^{common}$,
the product of all individual efficiencies for all benchmark points
and for the two background types.

\begin{table}
\caption{Selection criteria efficiencies (\%) for the background and signal
benchmark points. The efficiency of each criterion is listed after
all the previous ones have been applied. \label{tab:efficiencies}}

\smallskip{}

\begin{centering}
\begin{tabular}{c|c|c|c|c|c}
selection criterion  & S1  & S2  & S3  & $2\mu\,4j$ background  & $t\bar{t}$ background\tabularnewline
\hline
\hline 
at least 2$\mu$  & 63.6  & 77.9  & 84.1  & 93.3  & 8.1\tabularnewline
\hline 
$p_{T}(\mu)$>15 GeV  & 50.7  & 55.1  & 95.1  & 88.8  & 29.5\tabularnewline
\hline 
at least 4$j$  & 73.6  & 82.3  & 82.6  & 86.0  & 88.7\tabularnewline
\hline 
$p_{T}(j)>$15 GeV  & 53.3  & 65.6  & 72.2  & 70.4  & 76.0\tabularnewline
\hline 
$|M_{jj}-M_{W}|<20$ GeV  & 63.1  & 60.5  & 60.3  & 45.9  & 52.8\tabularnewline
\hline 
$\Delta R_{\mu j}>0.4$  & 64.5  & 65.9  & 77.4  & 83.0  & 17.4\tabularnewline
\hline 
no $j_{b}$  & 93.6  & 92.0  & 91.5  & 93.6  & 53.4\tabularnewline
\hline 
$E_{T}^{miss}<30$ GeV  & 74.4  & 64.9  & 68.7  & 79.4  & 15.4\tabularnewline
\hline
\hline 
$\epsilon_{reco}^{common}$  & 3.7  & 5.7  & 13.4  & 24.2  & $1.2\times10^{-2}$\tabularnewline
\end{tabular}
\par\end{centering}
\end{table}

\subsection{Dirac vs Majorana Neutrinos}

The presented event selection and reconstruction should be extended
depending on the Dirac or Majorana nature of the fourth family neutrinos.
In the Dirac case, the fourth family neutrinos and their anti-particles
are distinct; therefore the muons in the final state are expected
to be of opposite sign. In the Majorana case, however, 50\% of the
time the muons in the final state is of the same sign. The following
analysis deals with Dirac and Majorana cases separately.

\subsubsection{Majorana Case}

\begin{figure}
\begin{centering}
\includegraphics[scale=0.56]{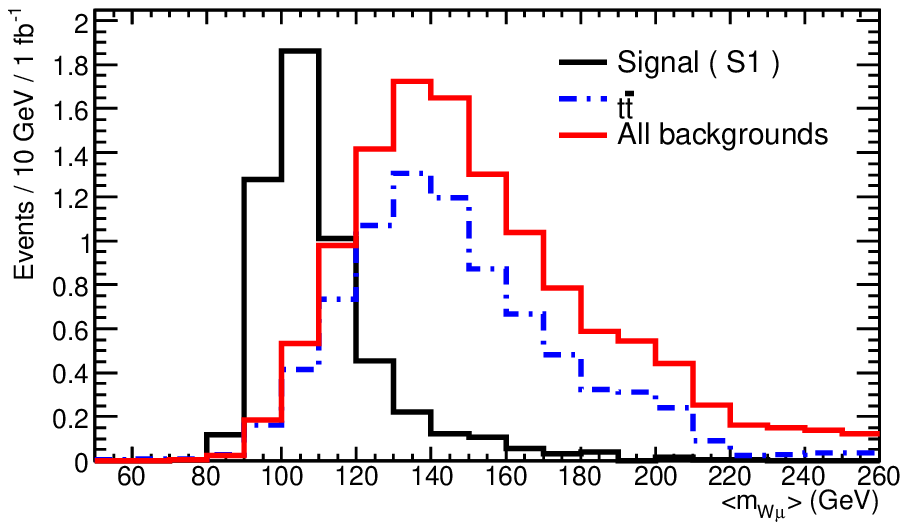}\includegraphics[scale=0.56]{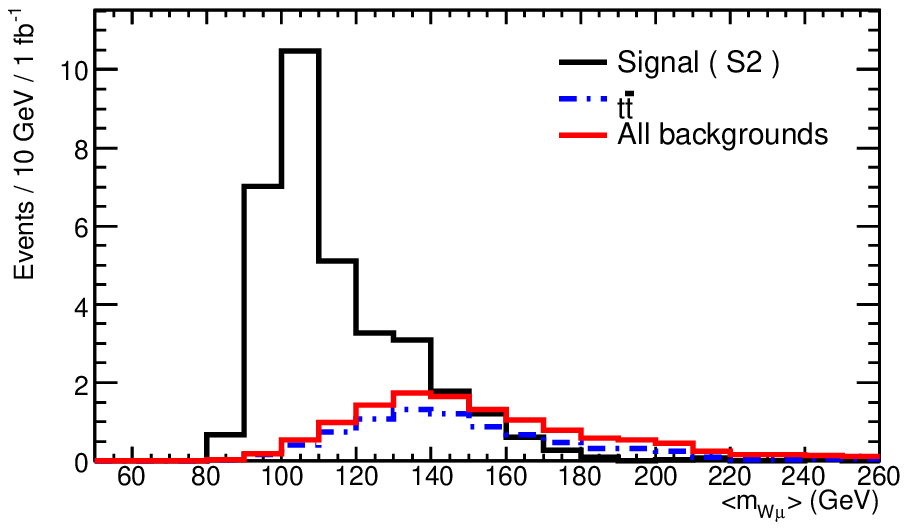}\includegraphics[scale=0.56]{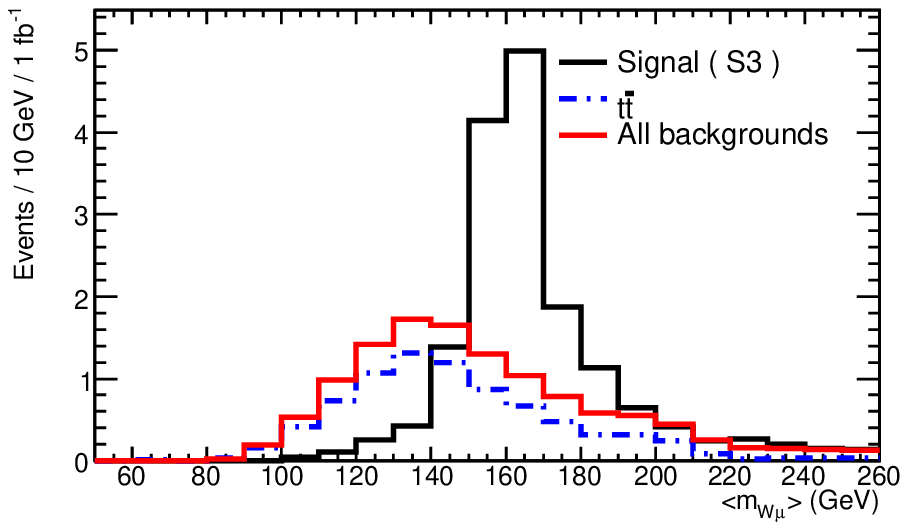}
\par\end{centering}

\begin{centering}
\includegraphics[scale=0.86]{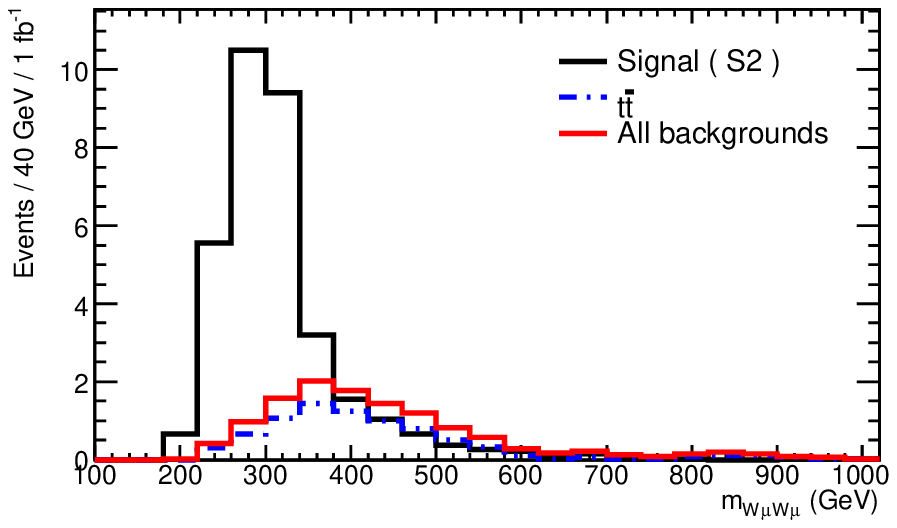}\includegraphics[scale=0.86]{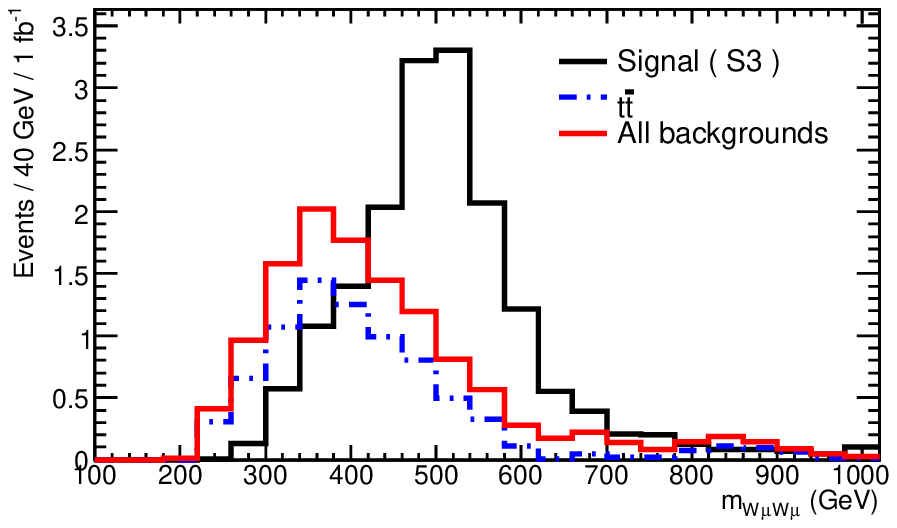} 
\par\end{centering}

\caption{\underbar{Majorana case:} Expected event yields for the three benchmark
points S1, S2 and S3 (from left to right). Histograms on the upper
row show the average of the invariant masses of the two $\nu_{4}$
candidates from each event, and the lower row shows the invariant
masses of the reconstructed Higgs boson candidates. In all plots,
the signal and background events are shown by solid black and solid
gray lines, respectively. The $t\bar{t}$ component of the background
is represented by the dashed histogram.\label{fig:majo-events}}

\end{figure}

The requirement of having same sign muons largely eliminates the SM
backgrounds as seen in table \ref{tab:final-majorana}. To further
eliminate the background events, the ratio of the mass difference
between the two $\nu_{4}$ candidates and their average is required
to be less than 0.25. Although the last requirement ensures a consistent
reconstruction of both $\nu_{4}$ candidates, only their average is
shown in the final invariant mass histograms in Fig.~\ref{fig:majo-events}
upper row for 1 fb$^{-1}$ of integrated luminosity. The lower two
plots in the same figure show the invariant mass distribution in the
$s$-channel for the two benchmark points with $m_{h}=300$ and $m_{h}=500$
GeV. In all plots, the signal is observed to be well above the background.

\begin{table}
\caption{Additional selection criteria for Majorana type fourth family neutrinos.}

\smallskip{}

\begin{centering}
\begin{tabular}{c|c|c|c|c|c}
selection criterion  & S1  & S2  & S3  & $2\mu\,4j$ background  & $t\bar{t}$ background\tabularnewline
\hline
\hline 
Sign($\mu_{1}$)$\times$Sign($\mu_{2}$) =1  & 46.6  & 45.5  & 51.2  & $6.8\times10^{-2}$  & 15.5\tabularnewline
\hline 
$\Delta M_{\nu_{4}}^{reco}\,/\,\overline{M_{\nu_{4}}^{reco}}$ < 0.25  & 88.2  & 85.0  & 74.3  & 52.0  & 58.8\tabularnewline
\hline 
$\epsilon_{total}^{MAJORANA}$  & 1.5 & 2.1  & 5.3  & $8.6\times10^{-3}$  & $1.1\times10^{-3}$\tabularnewline
\end{tabular}\label{tab:final-majorana} 
\par\end{centering}
\end{table}

\subsubsection{Dirac Case}

The charge requirement on the muons does not reduce the background
as heavily as in the Majorana case as shown in the first line of table
\ref{tab:final-dirac}. To further eliminate the background events,
the di-muon invariant mass (shown on the upper right-hand of in Fig.~\ref{fig:S1-kinematics})
is required to be at least 25$\,$GeV away from the nominal mass of
the $Z$ boson. Furthermore, to reject the muon pairs from the $\gamma_{\mu\mu}^{*}+4j$
events and from the cascade decays of $b$ quarks, an angular separation
$\Delta R\geq$2.0 is required. The $\Delta R\equiv\sqrt{(\Delta\eta)^{2}+(\Delta\phi)^{2}}$
distributions for the signal and two types of background are shown
in the upper left-hand plot in Fig.~\ref{fig:S1-kinematics}.

\begin{figure}
\begin{centering}
\includegraphics[width=0.46\textwidth]{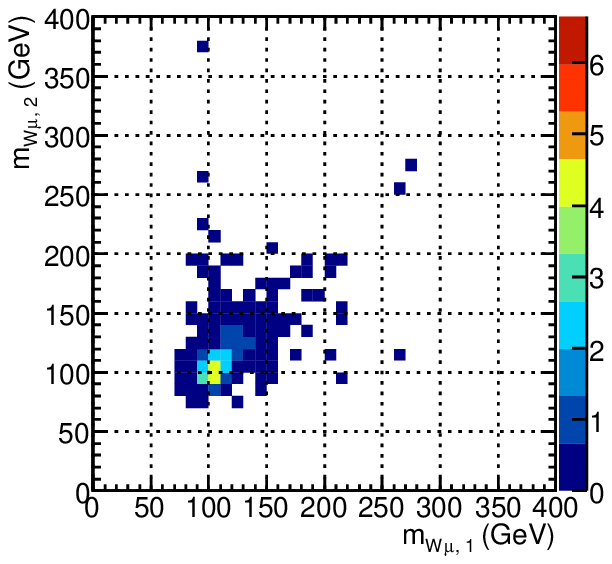}\includegraphics[width=0.46\textwidth]{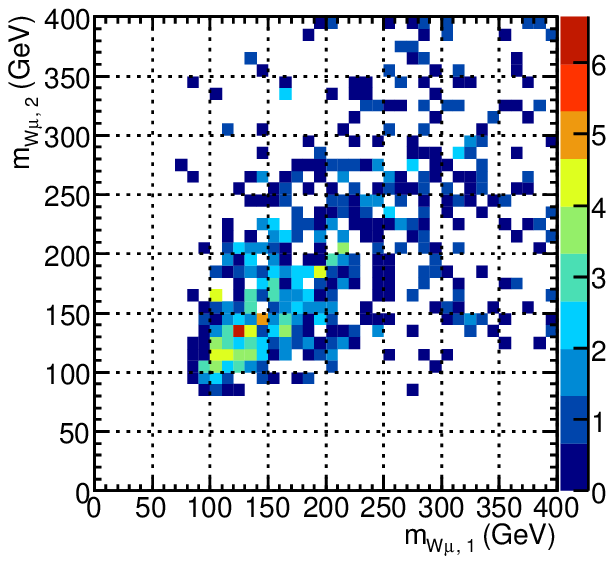} 
\par\end{centering}

\caption{The invariant masses of the two reconstructed $\nu_{4}$ candidates
for the $S2$ signal (left) and the sum of all backgrounds (right).}

\label{fig:TwoDcutExample} 
\end{figure}

\begin{figure}
\begin{centering}
\includegraphics[scale=0.56]{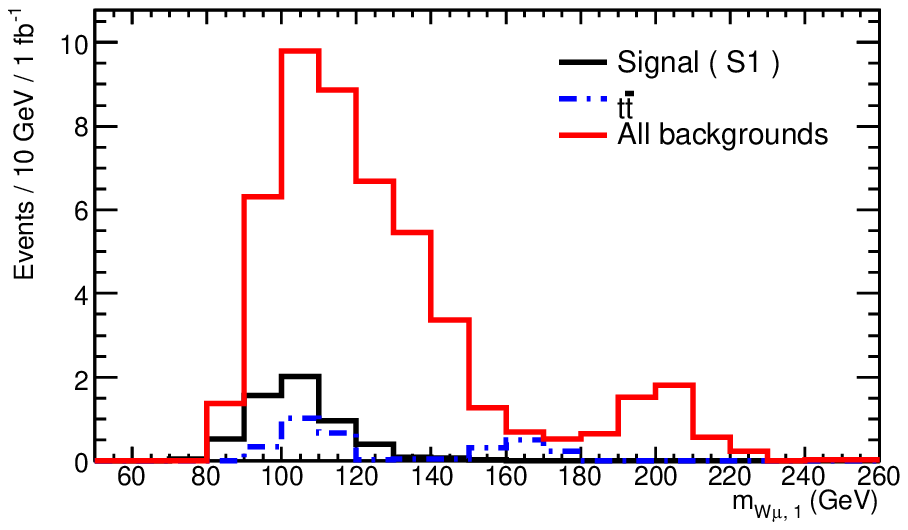}\includegraphics[scale=0.56]{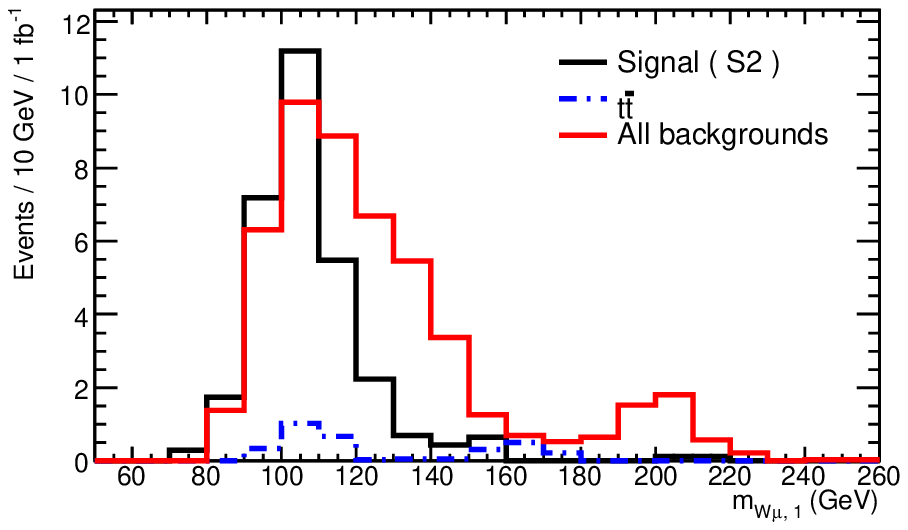}\includegraphics[scale=0.56]{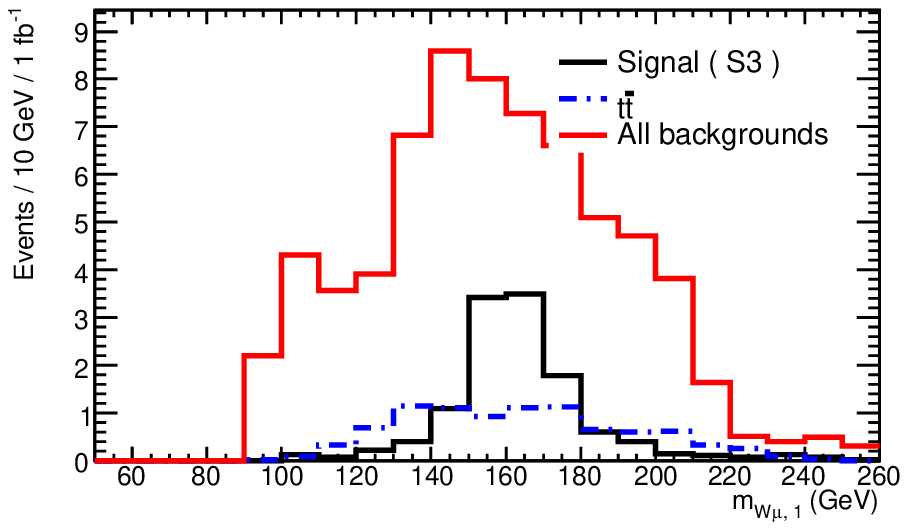} 
\par\end{centering}

\begin{centering}
\includegraphics[scale=0.86]{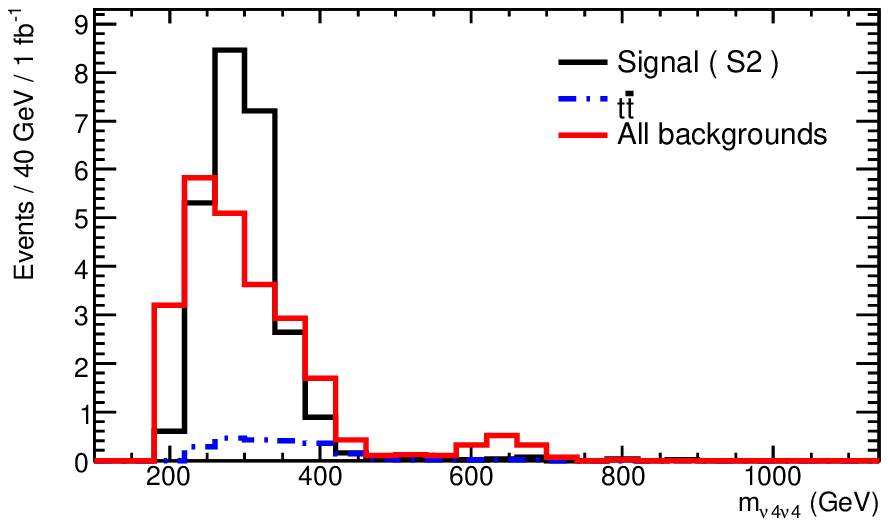}\includegraphics[scale=0.86]{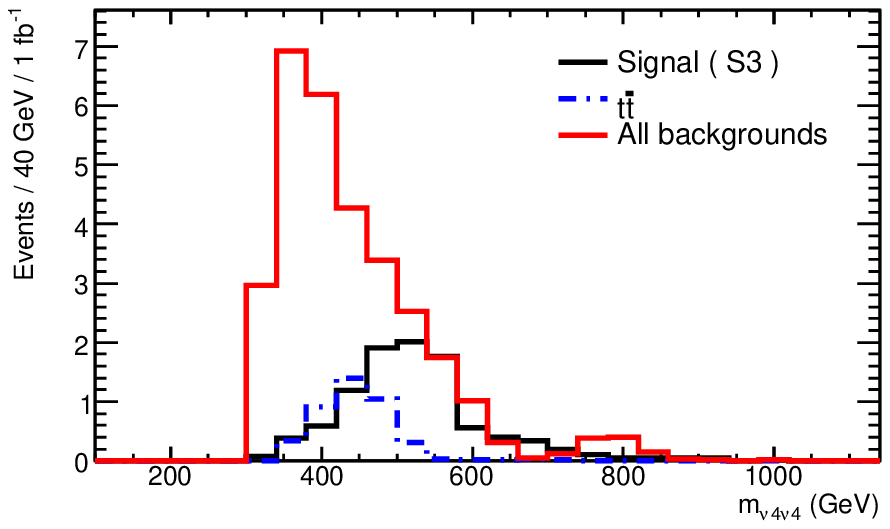} 
\par\end{centering}

\caption{\underbar{Dirac case:} Expected event yields for the three benchmark
points S1, S2 and S3 (from left to right). Histograms on the upper
row show the invariant mass of one of the two $\nu_{4}$ candidates
from each event, when the other candidate is required to be within
20$\,$GeV of the true mass. The lower row shows the invariant masses
of the reconstructed Higgs boson candidates when both $\nu_{4}$ candidates
satisfy the sliding window cuts. In all plots, the signal and background
events are shown by solid black and solid gray lines, respectively.
The $t\bar{t}$ component of the background is represented by the
dashed histogram.\label{fig:dirac-events}}

\end{figure}

Regardless of the considerable reduction obtained from these cuts,
the $2\mu\,4j$ background is still quite significant. Therefore a
two dimensional selection window of $m\pm$20$\,$GeV is considered
in the $m_{\nu_{4}1}^{reco}-m_{\nu_{4}2}^{reco}$ plane (Fig.~\ref{fig:TwoDcutExample}).
With the actual data, the centre point of this {}``sliding'' window
can be moved to search for an excess of events. For this feasibility
study, the sliding selection box is centered around the true value
of the $\nu_{4}$ mass ($m=m_{\nu_{4}}^{true}$ ). The selection efficiency
for this two dimensional selection criteria as well as the final total
efficiencies for all benchmark points are listed in table \ref{tab:final-dirac}.
The invariant mass of one of the two $\nu_{4}$ candidates, when the
other is within the sliding window, and the reconstructed Higgs boson
invariant mass, when both $\nu_{4}$ candidates are in the sliding
window, can be found in Fig.~\ref{fig:dirac-events} in the upper
and lower rows, respectively. 

\begin{table}
\caption{Additional selection criteria for Dirac type fourth family neutrinos.}

\smallskip{}

\begin{centering}
\begin{tabular}{c|c|c|c|c|c|c|c}
selection criterion  & S1  & S2  & $2\mu\,4j$ background  & $t\bar{t}$ background  & S3  & $2\mu\,4j$ background  & $t\bar{t}$ background\tabularnewline
\hline
\hline 
Sign($\mu_{1}$)$\times$Sign($\mu_{2}$) =-1  & 97.3  & 96.5  & 99.9  & 84.5  & 99.2  & 99.9  & 84.5\tabularnewline
\hline 
$|m_{\mu^{+}\mu^{-}}-m_{Z}|>25$ GeV & 79.1 & 74.1 & 10.0 & 67.7 & 77.6 & 10.0 & 67.7\tabularnewline
\hline 
$\Delta R_{\mu\mu}>2.0$  & 72.9 & 65.6 & 34.3 & 59.5 & 74.7 & 34.3 & 59.5\tabularnewline
\hline 
$|m_{\nu_{4}}^{reco}-m_{\nu_{4}}^{true}|<20$ GeV  & 67.9  & 60.4  & 5.5  & 6.1  & 39.6  & 6.06  & 13.6\tabularnewline
\hline 
$\epsilon_{total}^{DIRAC}$  & 1.4  & 1.6  & $4.5\times10^{-2}$  & $2.5\times10^{-4}$  & 3.1  & $5.0\times10^{-2}$ & $5.7\times10^{-4}$\tabularnewline
\end{tabular}\label{tab:final-dirac} 
\par\end{centering}
\end{table}

\section{Results and Discussion}

The number of expected signal ($s$) and background ($b$) events
are obtained by integrating the contents of the 4 bins around the
signal peak for each of the histograms shown in Figures \ref{fig:majo-events}
and \ref{fig:dirac-events}. For the $\nu_{4}$ histograms in the
Dirac case, this exactly corresponds to counting the number of events
in the sliding window. The statistical significance of the expected
signal was calculated using the definition: ${\cal S}=\sqrt{2\times[(s+b)\ln(1+\frac{s}{b})-s]}\;$$~$\cite{CMS_significance}.
The table \ref{tab:generic_results}, contains the number of signal
and background events, and the significance for the three benchmark
points. The signal significance as a function of the integrated luminosity
is given in Fig.\ref{fig:generic_reach} for both $\nu_{4}$ and $h$
signals. For non-zero significance, a minimum of 3 signal events were
required. It is seen that, depending on their masses, an early double
discovery of both the Higgs boson and the fourth family neutrino is
possible in the first year of the LHC operation, i.e., with one fb$^{-1}$
of data. Furthermore, the integrated luminosity necessary to claim
a 3$\sigma$ observation and a 5$\sigma$ discovery is given in table
\ref{tab:dirac-majorana_results} for the all considered scenarios.

\begin{center}
\begin{table}
\caption{For the three benchmark points, the statistical significance for the
discovery of the heavy neutrino $\nu_{4}$ and of the Higgs boson
estimated at 1fb$^{-1}$ of integrated luminosity.}

\begin{centering}
\smallskip{}
 \label{tab:generic_results}\begin{tabular}{l|c|c|c||c|c|c}
Benchmark  & \multicolumn{1}{c}{} & \multicolumn{1}{c}{$\nu_{4}$} & \multicolumn{1}{c||}{} & \multicolumn{1}{c}{} & \multicolumn{1}{c}{$h$} & \multicolumn{1}{c}{}\tabularnewline
Point  & signal  & backgrd  & significance  & signal  & backgrd  & significance\tabularnewline
\hline
\hline 
\emph{S1$_{Dirac}$}  & 5.1  & 26.3  & 1.0  & N/A  & N/A  & N/A\tabularnewline
\hline 
\emph{S2$_{Dirac}$}  & 25.6  & 26.3  & 4.4  & 23.6  & 17.5  & 4.8\tabularnewline
\hline 
\emph{S3$_{Dirac}$}  & 9.8  & 30.5  & 1.7  & 6.9  & 11.9  & 1.8\tabularnewline
\hline
\hline 
\emph{S1$_{Majorana}$}  & 4.2  & 1.4  & 2.7  & N/A  & N/A  & N/A\tabularnewline
\hline 
\emph{S2$_{Majorana}$}  & 23.2  & 1.4  & 9.7  & 28.7  & 5.0  & 8.4\tabularnewline
\hline 
\emph{S3$_{Majorana}$}  & 12.4  & 4.7  & 4.4  & 10.6  & 4.0  & 4.1\tabularnewline
\end{tabular}
\par\end{centering}
\end{table}

\par\end{center}

\begin{center}
\begin{figure}
\begin{centering}
\includegraphics[scale=0.45]{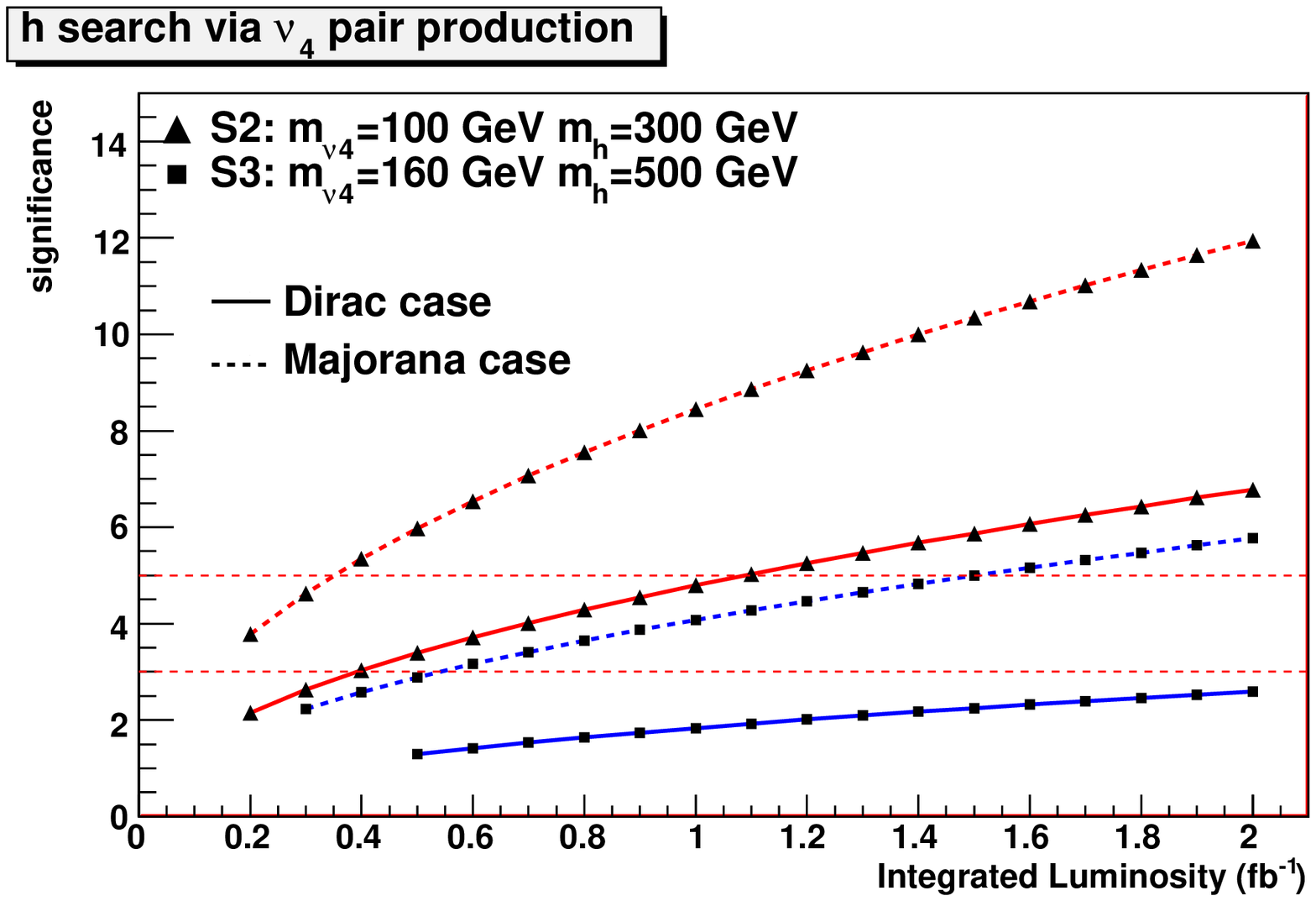}\includegraphics[scale=0.45]{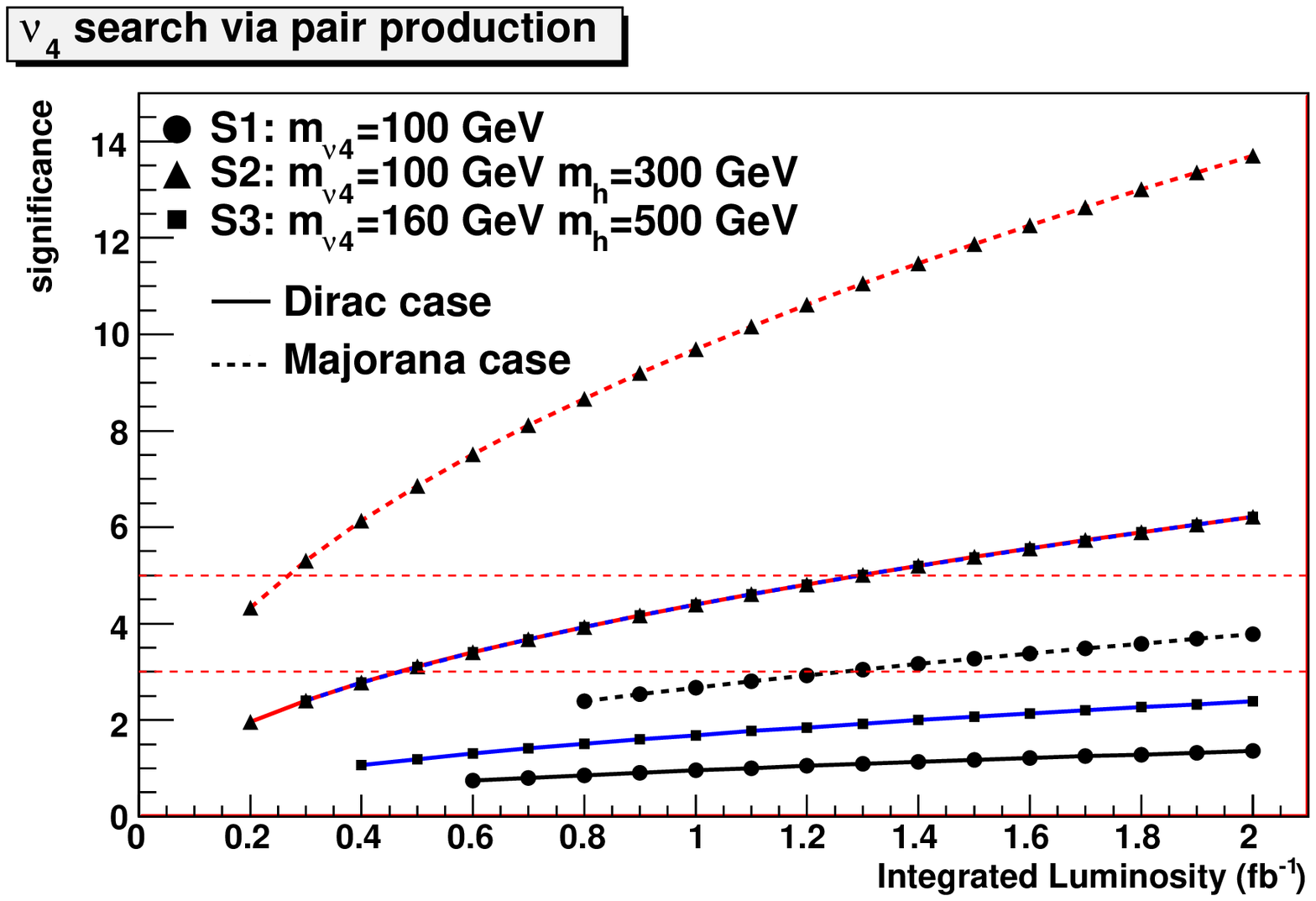} 
\par\end{centering}

\caption{Expected signal significance for the Higgs boson and fourth family
neutrino searches. For each point on the curves, at least 3 signal
events are required to have satisfied all the selection criteria.\label{fig:generic_reach}}

\end{figure}

\par\end{center}

\begin{table}
\caption{Required integrated luminosity in pb$^{-1}$ for 3 (5) $\sigma$ statistical
significance for the discovery of the heavy neutrino $\nu_{4}$ and
of the Higgs boson, for the three benchmark points.}

\begin{centering}
\smallskip{}
 \label{tab:dirac-majorana_results}\begin{tabular}{l|c|c||c|c}
Benchmark  & \multicolumn{2}{c}{$\nu_{4}$} & \multicolumn{2}{c}{$h$}\tabularnewline
Point  & Dirac  & Majorana  & Dirac  & Majorana\tabularnewline
\hline
\hline 
\emph{S1}  & 9800 (27000)  & 1300 (3500)  & N/A  & N/A\tabularnewline
\hline 
\emph{S2}  & 470 (1300)  & 100 (270)  & 390 (1100)  & 130 (350)\tabularnewline
\hline 
\emph{S3}  & 3200 (8800)  & 470 (1300)  & 2700 (7400)  & 540 (1500)\tabularnewline
\end{tabular}
\par\end{centering}
\end{table}

\section{Conclusions\label{sec:Conclusions}}

While hadron colliders are not considered to be the best place to
search for heavy charged and neutral leptons due to small production
cross section, the existence of the Higgs particle might drastically
change this picture. For example, if the Higgs mechanism is the one
that Nature choose to give masses to the fermions, the LHC has the
chance to simultaneously discover both the Higgs boson itself and
the fourth family neutrino using the $pp\to h\to\nu_{4}\bar{v}_{4}$
channel. The main reason for this possibility is the enhancement of
the gluon fusion process due to fourth family quarks yielding a high
Higgs boson production rate. If $m_{h}=300$$\,$GeV and $m_{\nu_{4}}=100$$\,$GeV,
LHC would discover both of them with 5$\sigma$ significance even
with an integrated luminosity of around 350 pb$^{-1}$ , provided
the fourth family neutrinos are of Majorana nature. Alternatively,
if they are of Dirac nature, the double discovery of $\nu_{4}$ and
$h$ is again possible with less than 1.5fb$^{-1}$ of data. For heavier
particles ($m_{h}=500$ GeV and $m_{\nu_{4}}=160$ GeV), the signals
from Majorana (Dirac) type neutrinos and the Higgs boson can also
be observed with about 1.5 (9) fb$^{-1}$ of data. A similar result
has been obtained for another process ($pp\to W^{+}\to\ell^{+}N\to\ell^{+}\ell^{+}jj$)
in a recent paper$\,$\cite{single-v4}.

Finally, if the Higgs boson does not exist, the $Z$ boson provides
the only tree-level channel for the pair production of fourth family
neutrinos. This study shows that the 5$\sigma$ significance can be
attained with about 3.5 fb$^{-1}$ for the Majorana type and about
30 fb$^{-1}$ for Dirac type neutrinos. However, a more detailed search,
also involving the semi-leptonic di-$W$ boson decays would reduce
the amount of data-taking time needed. Such a study is in progress.

\subsection*{Acknowledgments}

The authors would like thank Paul Coe and Jeff Tseng for useful suggestions.
S.S. acknowledges the support from the Turkish State Planning Committee
under the contract DPT2006K-120470 and from the Turkish Atomic Energy
Authority (TAEK). G.Ü.'s work is supported in part by U.S. Department
of Energy Grant DE FG0291ER40679. M.K.Ü., T.D. and V.E.Ö. acknowledge
support from the UK Science and Technology Facilities Council.

\end{document}